\documentclass[
 reprint,
 superscriptaddress,
 amsmath,amssymb,
 aps,
 prd
]{revtex4-2}

\usepackage[T1,T2A]{fontenc}
\usepackage[unicode,final]{hyperref}
\usepackage{amsmath}
\usepackage{amssymb}
\usepackage{amsthm}
\usepackage{orcidlink}
\usepackage{booktabs}
\usepackage{multirow}
\usepackage{mleftright}
\usepackage{colortbl}
\usepackage{subcaption}
\usepackage{cleveref}
\newcommand\grey[1]{\multicolumn1{>{\columncolor{gray!50}}c}{#1}}
\newcommand\black[1]{\multicolumn{#1}{>{\columncolor{black!80}}c}{}}
\newcommand\blackwith[1]{\multicolumn1{>{\columncolor{black!80}}c}{\color{white}#1}}

\usepackage[british]{babel}
\usepackage[artemisia]{textgreek}
\usepackage[utf8]{inputenc}
\usepackage{tikz-cd}
\usetikzlibrary{babel}

\usepackage[textsize=tiny]{todonotes}

\newcommand\coloneqq{\mathrel{:=}}
\newcommand\eqqcolon{\mathrel{=:}}

\begin{document}

\preprint{In preparation for submission to PRD}

\title{Sandwich Construction of Symmetry TFTs for
the Centre Symmetries of Chern–Simons,
Yang–Mills, and Einstein Gravity}



\author{Leron  Borsten}
 \email{l.borsten@herts.ac.uk}
\affiliation{%
Department of Physics, Astronomy and Mathematics\\University of Hertfordshire, Hatfield, Herts.\ Al10 9AB, United Kingdom
}%
\affiliation
 {Blackett Laboratory, Imperial College London, London SW7 2AZ, United Kingdom
}

\author{Dimitri  Kanakaris}%
 \email{d.kanakaris-decavel@herts.ac.uk}
\affiliation{%
Department of Physics, Astronomy and Mathematics\\University of Hertfordshire, Hatfield, Herts.\ Al10 9AB, United Kingdom
}%

\author{Hyungrok Kim}
 \email{h.kim2@herts.ac.uk}
\affiliation{%
Department of Physics, Astronomy and Mathematics\\University of Hertfordshire, Hatfield, Herts.\ Al10 9AB, United Kingdom
}



\begin{abstract}
	We construct symmetry topological field theories (SymTFTs) using the sandwich construction of Pulmann-Ševera-Valach that manifest the centre symmetries of Chern-Simons theory and Yang-Mills theory as well as general relativity in the MacDowell-Mansouri formulation.
The `filling' of the sandwich is an AKSZ sigma model whose target space is a Weil algebra, augmented with discrete degrees of freedom given by a choice of topological boundary condition.
\end{abstract}


\maketitle

\tableofcontents

\section{Introduction and Summary}

The seminal work of \cite{Gaiotto:2014kfa}  vastly generalised  our   notion symmetry in quantum field theory by shifting to the perspective that a global symmetry corresponds to a topological
defect operator, as reviewed in  \cite{Cordova:2022ruw,Brennan:2023mmt,Gomes:2023ahz,Bhardwaj:2023kri,Luo:2023ive,Schafer-Nameki:2023jdn}. This has revealed a vast menagerie of \emph{generalised} or \emph{categorical} symmetries (which need not be invertible), going far beyond the traditional group theoretic conception, with important physical implications  and applications. 
An important class of such generalised  symmetries are the global $p$-form symmetries (also referred to as `higher-form symmetries'), which  abound in quantum field theory  and have a long history \cite{Polyakov:1976fu,Polyakov:1975rs,tHooft:1977nqb,Kovner:1992pu,deWildPropitius:1995hk,Alford:1990fc, Bucher:1991bc, Alford:1992yx, Pantev:2005rh,Pantev:2005zs,Hellerman:2006zs,Nussinov:2009zz}. In particular, they play a crucial role in the modern understanding of topological phases \cite{Gukov:2013zka,Kapustin:2013uxa,Barkeshli:2014cna,Gukov:2014gja,Nguyen:2024ikq}.

More recently, it has been discovered 
that such generalised symmetries are organised and classified by a \emph{symmetry topological field theory} (SymTFT) \cite{Apruzzi:2021nmk,Bhardwaj:2023ayw, Bhardwaj:2023fca},
which is a topological field theory in one higher dimension that, under appropriate boundary conditions,
dimensionally reduces to the non-topological theory whose symmetries we are considering,
as reviewed in \cite{Bhardwaj:2023kri,Schafer-Nameki:2023jdn}.
Many applications and constructions of SymTFTs are now known, and they remain a topic of intense investigation \cite{Ji:2019jhk,Gaiotto:2020iye,Apruzzi:2021nmk,Apruzzi:2022rei,Chatterjee:2022tyg,Kaidi:2022cpf,Antinucci:2022vyk,Kaidi:2023maf,vanBeest:2022fss,Bhardwaj:2023ayw,Bhardwaj:2023bbf,Gagliano:2024off,Bhardwaj:2024igy,Antinucci:2024ltv,Jia:2025jmn,Schafer-Nameki:2025fiy,Jia:2025bui,Antinucci:2025fjp}.

In particular, \(d\)-dimensional Yang-Mills theory with gauge group \(G\) enjoys a \(\operatorname Z(G)\)-valued one-form electric symmetry as well as a \(\pi_1(G)\)-valued \((d-3)\)-form magnetic symmetry \cite{Polyakov:1976fu,Polyakov:1975rs,tHooft:1977nqb}, which plays a crucial role in confinement (as reviewed in \cite{Holland:2000uj,Ogilvie:2012is}; cf.\ the recent discussions in \cite{Nguyen:2024ikq,Hayashi:2024yjc,Giansiracusa:2025hfj, Borsten:2025diy}).
One can trade the electric symmetry for the magnetic one or vice versa by gauging the respective discrete symmetries, and this corresponds to changes in topological boundary conditions of the SymTFT.
Relatedly, recent works have investigated generalised symmetries in gravity \cite{Hinterbichler:2022agn, Gomez-Fayren:2023qly,Hull:2024bcl,Hull:2024ism,Cheung:2024ypq,Hull:2025ivk}, and in particular Einstein gravity enjoys a one-form symmetry valued in the centre of the Lorentz group similar to Yang-Mills theory \cite{Cheung:2024ypq}.

There are many ways to construct such SymTFTs, cf.~\cite{Apruzzi:2021nmk,vanBeest:2022fss,Apruzzi:2023uma,Bhardwaj:2023bbf,Gagliano:2024off,Tian:2024dgl,Jia:2025jmn}, including top-down constructions from string/M-theory.
In this paper, we provide a bottom-up construction for the centre symmetries of Yang-Mills theory and gravity.
The sandwich construction for SymTFTs originates in \cite{Severa:2016prq,Pulmann:2019vrw} (see also \cite{Freed:2022qnc,Freed:2022iao}),
where the \((d+1)\)-dimensional bulk theory (the `filling' of the sandwich) belongs to a family of topological field theories known as the Alexandrov-Kontsevich-Schwarz-Zaboronsky (AKSZ) model \cite{Alexandrov:1995kv} (reviewed in \cite{Ikeda:2012pv,Roytenberg:2006qz}). 
From this perspective,  the powerful apparatus of shifted symplectic structures \cite{Pantev:2013fov,calaque14,Calaque:2013xja,Calaque:2021sgp} and symplectic differential graded geometry provides an elegant account of the dimensional reduction to the \(d\)-dimensional boundary theory \cite{Pulmann:2019vrw}.
The resulting \(d\)-dimensional boundary theory is often naturally of the form of a \(d\)-dimensional AKSZ model perturbed by a non-topological quadratic term;
this family of theories is known as Manin-type theories and appears naturally in many contexts \cite{Arvanitakis:2024dbu,Borsten:2024pfz,Borsten:2024alh,Arvanitakis:2025nyy} and includes Yang-Mills theory \cite{Arvanitakis:2024dbu,Arvanitakis:2025nyy}, two-, three-, and four-dimensional gravitational models \cite{Borsten:2024pfz,Borsten:2024alh}, the third-way model \cite{Arvanitakis:2024dbu}, and integrable sigma models \cite{Arvanitakis:2025nyy}.
It is therefore natural to directly engineer SymTFTs for Manin-type theories such as Einstein gravity and Yang-Mills theory using the language of symplectic differential graded geometry, and it is this construction that we explain in the present contribution.

This paper is organised as follows. \Cref{sec:review} reviews the AKSZ construction, the Pulmann-Ševera-Valach sandwich construction of SymTFTs, and required elements of differential graded geometry and stacks. \Cref{sec:triv-filling} then explains how one can realise an AKSZ model as the dimensional reduction of a trivial higher-dimensional theory.
Adding in stacky discrete degrees of freedom, \cref{sec:cs} then explains how to construct the centre symmetry of three-dimensional Chern-Simons theory as the simplest example of an AKSZ model.
Perturbing the AKSZ construction with Manin terms, \cref{sec:ym} and \cref{sec:gr} then construct sandwich SymTFTs that capture the centre symmetries of Yang-Mills theory in any spacetime dimension and Einstein gravity in four dimensions respectively.

\section{Review  of  the sandwich recipe}\label{sec:review}
We first briefly review the mathematical formalism behind the Pulmann-Ševera-Valach sandwich construction.

\subsection{Graded geometry}
Due to the complicated gauge symmetries involved, it is simplest to work within the formalism of graded geometry and the Batalin-Vilkovisky formalism (see \cite{Cattaneo:2010re,Qiu:2011qr} for reviews of this formalism).

The main geometric objects we work with are \emph{differential graded manifolds} \(M\), which are `smooth manifolds' whose coordinates carry integer degrees and whose ring of functions \(\mathcal C^\infty(M)\) carries a differential
\begin{equation}
    Q\colon\mathcal C^\infty(M)\to\mathcal C^\infty(M)
\end{equation}
of degree \(1\) that obeys the graded Leibniz rule with respect to multiplication:
\begin{equation}
    Q(fg) = (Qf)g + (-1)^{|f|}f(Qg)
\end{equation}
for \(f,g\in\mathcal C^\infty(M)\), where \(|f|\in\mathbb Z\) denotes the degree of a homogeneous element.
Since \(Q\) obeys the Leibniz rule, it can be regarded as a vector field of degree \(1\) on \(M\) and is thus called the \emph{homological vector field}.

The usual ingredients of differential geometry, such as vector fields and differential forms, work in the same way as before except that now objects must carry grading. In the case of differential forms on a graded manifold, now objects carry two gradings: the form degree and the intrinsic degree of the coordinates. A \(p\)-form \(\alpha\) of intrinsic degree \(q\) and a \(p'\)-form \(\alpha'\) of intrinsic degree \(q'\) obey the bigraded commutation relation
\begin{equation}
    \alpha\wedge\alpha'
    =(-1)^{pp'+qq'}\alpha'\wedge\alpha.
\end{equation}
For more on the issues of grading, see \cite{zbMATH01735158}.

To shift degrees of graded objects, we use the suspension notation \((-)[i]\), so that if \(V=\bigoplus_iV_i\) has components \(V_i\) of degree \(i\), then the shifted object \(V[i]\) has components with degrees
\begin{equation}
    (V[n])_j
    =V_{n+j}.
\end{equation}
So, for example, if \(V=V_i\) is concentrated in degree \(i\), then \(V[n]\) is concentrated in degree \(j=i-n\).

\subsubsection{Maps between differential graded manifolds}
Between two differential graded manifolds \(X\) and \(Y\), one can consider three classes of maps, depending on how much structure they preserve:
\begin{itemize}
\item the class \(\operatorname{Maps}(X,Y)\) of \emph{all} smooth maps \(X\to Y\) that do not necessarily preserve the degrees of the coordinates;
\item the class of smooth maps \(\operatorname{Maps_g}(X,Y)\) that preserve the degrees of the coordinates;
\item the class of smooth maps \(\operatorname{Maps_{dg}}(X,Y)\) that preserve the degrees of the coordinates as well as the homological vector field.
\end{itemize}
For the purposes of the AKSZ formalism, it is most convenient to work with the first class, \(\operatorname{Maps}(X,Y)\),  most of the time. For example, let \(U\) be a contractible manifold, and let \(\mathfrak g\) be a Lie algebra. Then both \(\mathrm T[1]U\) and \(\mathfrak g[1]\) canonically carry the structure of differential graded manifolds, and the three classes of maps between them are characterised as follows:
\begin{itemize}
\item \(\operatorname{Maps}(\mathrm T[1]U,\mathfrak g[1])=\Omega(U)\otimes\mathfrak g[1]\) is the space of all \(\mathfrak g\)-valued differential forms on \(U\), or equivalently (when \(U\) is three-dimensional) the space of fields of Chern-Simons theory on \(U\) in the Batalin-Vilkovisky formalism including the ghosts and the antifields.
\item \(\operatorname{Maps_g}(\mathrm T[1]U,\mathfrak g[1])=\Omega^1(U)\otimes\mathfrak g[1]\) is the space of \(\mathfrak g\)-valued one-forms on \(U\), or equivalently the space of all \(\mathfrak g\)-valued connections on \(U\).
\item \(\operatorname{Maps_{dg}}(\mathrm T[1]U,\mathfrak g[1])=\{A\in\Omega^1(U)\otimes\mathfrak g[1]~|~\mathrm dA+\frac12[A,A]=0\}\) is the space of flat \(\mathfrak g\)-valued connections on \(U\).
\end{itemize}

\subsubsection{Linearisation to \(L_\infty\)-algebras}
Given a point \(x\in X\) in an ordinary manifold \(X\), then the tangent space \(\mathrm T_xX\), which is a real vector space can be thought of as a linearisation of \(X\) at \(x\); it is a vector space which is the `closest approximation' to \(X\) at that point. When \(X\) is a graded manifold, then the tangent space \(\mathrm T_xX\) carries more structure: it is a graded real vector space. If \(X\) is furthermore a differential graded manifold, then the shifted tangent space \(\mathrm T[-1]_xX\) carries the structure of a curved \(L_\infty\)-algebra \cite{Jurco:2018sby}.
Finally, if \((X,\omega_X)\) is a symplectic differential graded manifold with \(|\omega_X|=n\), then the tangent space \(\mathrm T[-1]_xX\) forms a cyclic \(L_\infty\)-algebra with cyclic structure of degree \(n-2\) \cite{Jurco:2018sby}.

\subsubsection{Differential graded symplectic structure}
A \emph{symplectic structure} on a differential graded manifold \((X,Q_X)\) is a two-form \(\omega_X\) on \(X\) that is closed and nondegenerate and furthermore is invariant under the Lie derivative by \(Q_X\):
\begin{equation}
    \mathcal L_{Q_X}\omega_X = 0.
\end{equation}
The symplectic structure \(\omega_X\) can carry arbitrary intrinsic degree.
A symplectic differential graded manifold \((X,Q_X,\omega_X)\) is a differential graded manifold equipped with a symplectic structure.

A \emph{Lagrangian submanifold} of a symplectic differential graded manifold \((X,Q_X,\omega_X)\) is a homogeneous graded submanifold \(L\subset X\) that is maximally isotropic (i.e.\ \(\omega|_L = 0\) and \(L\) is maximal under inclusion amongst submanifolds satisfying this) and tangent to the homological vector field \(Q_X\).
Then the grading of \(X\) and the homological vector field \(Q_X\) restrict to \(L\), making \(L\) a differential graded manifold. (The symplectic structure does not descend to \(L\).)

A \emph{resolution} \cite[§6]{Pulmann:2019vrw} of a Lagrangian submanifold \(L\subset X\) is a better-behaved substitute for a Lagrangian submanifold: it is a differential graded smooth map \(l\colon R\to X\) together with a quasi-isomorphism of differential graded manifolds \(q\colon L\to R\) such that the composition
\begin{equation}
    L\xrightarrow q R\xrightarrow lX
\end{equation}
agrees with the inclusion \(L\hookrightarrow X\).

\subsection{AKSZ model}
The AKSZ model \cite{Alexandrov:1995kv} (reviewed in \cite{Ikeda:2012pv,Roytenberg:2006qz})
is a topological sigma model of the Schwarz type (i.e.\ explicitly independent of any metric on the source tangent space and not obtained by a topological twist) whose target space is a differential graded manifold \(X\) equipped with a symplectic form \(\omega_X\); if the intrinsic degree of \(\omega_X\) is \(n\), then the dimension of the worldvolume \(M\) is \(n+1\).
Special cases of the AKSZ model include the two-dimensional Poisson sigma model \cite{Schaller:1994es,Schaller:1994pm,Schaller:1994uj,Schaller:1995xk}, the three-dimensional Chern-Simons theory \cite{Witten:1988hf}, and the \(BF\) model \cite{Blau:1989bq,Blau:1989dh,Horowitz:1989ng,Myers:1989bp,Karlhede:1989hz} (reviewed in \cite{Birmingham:1991ty,Broda:2005wk}).
The AKSZ model serves as the topological `filling' of the SymTFT in in the Pulmann-Ševera-Valach sandwich construction.

Let \((X, Q_X,\omega_X)\) be a symplectic differential graded manifold with \(\deg\omega_X = n\). The target space is $X$.
The source is \(M\),  an oriented closed \((n+1)\)-dimensional smooth manifold (the worldvolume or spacetime).
Then the space of fields of the AKSZ model is the infinite-dimensional graded manifold
\begin{equation}
    \mathcal M \coloneqq \operatorname{Maps}(\mathrm T[1]M, X)
\end{equation}
of (graded) smooth maps that do \emph{not} necessarily preserve the homological vector field.

This space \(\mathcal M\) carries canonically the structure of a differential graded manifold with differential
\begin{equation}Q_{\mathcal M} = Q_X + Q_{\mathrm T[1]M}.\end{equation}
Furthermore, it carries the canonical symplectic form
\begin{equation}\omega_{\mathcal M}(u, v) =\int_{\mathrm T[1] M}\omega_X (u, v)\end{equation}
for tangent vectors \(u,v\in\mathrm T_f\mathcal M\) at some \(f\in\mathcal M\). The form \(\omega_{\mathcal M}\) has intrinsic degree \(1\), which makes it suitable as input to the Batalin-Vilkovisky formalism, which produces a classical action functional (called the \emph{akszion} in \cite{Pulmann:2019vrw}) that is manifestly topological (a metric on \(M\) has not been used) and gauge-invariant;
quantisation can then proceed as usual on the space of classical solutions to the variational equation of motion.

Note that the AKSZ action functional is gauge-invariant for small gauge transformations, but this is not automatic for large gauge transformations. For instance, for the special case of Chern--Simons theory with a compact simple gauge group, demanding invariance under large gauge transformation leads to quantisation of the level; fractional levels are inconsistent.

\subsection{AKSZ boundary conditions  as bread for the sandwich}
If \(\partial M\ne\varnothing\), \(\omega_{\mathcal M}\) is not \(Q_{\mathcal M}\)-invariant,
and one must impose suitable boundary conditions. In \cite{Pulmann:2019vrw} it is shown that an appropriate choice of boundary conditions is of the form
\[
    f|_{\partial M}\in\mathcal F
\]
for the fields \(f\in\mathcal M\), where \(\mathcal F\) is a Lagrangian submanifold of the space of boundary fields
\begin{equation}\mathcal X \coloneqq \operatorname{Maps}(\mathrm T[1]\partial M, X),\end{equation}
which is a differential graded manifold equipped with a canonical symplectic structure \(\omega_{\mathcal X}\) of intrinsic degree zero given by
\begin{equation}
    \omega_{\mathcal X}(u,v)\coloneqq\int_{\mathrm T[1]\partial M}\omega_X(u,v).
\end{equation}

In general, these boundary conditions \(\mathcal F\) can depend on geometric data (such as a Riemannian metric) on \(\partial M\) and thus partially break diffeomorphism symmetry.
One class of boundary conditions that is topological (in that does not depend on any additional geometric data on \(\partial M\) beyond the topology and the smooth structure) is given by the ansatz
\begin{equation}\label{eq:topological_ansatz}
    \mathcal F_L = \operatorname{Maps}(\mathrm T[1]\partial M, L),
\end{equation}
where \(L\) is a Lagrangian submanifold of the target space \(X\).

Note that, in a general SymTFT, one should consider \emph{all} possible topological boundary conditions, which classify the various (generalizsed) symmetries of the theory. The ansatz \eqref{eq:topological_ansatz} provides a wide range of topological boundary conditions, but it is not immediate that these exhaust the set of all possible such boundary conditions. (For a related discussion, see \cite{Pulmann:2019vrw}.)

\subsection{Sandwich construction and dimensional reduction}
We now specialise to the case where the \((d+1)\)-dimensional bulk is of the form \(M=\Sigma\times[0,1]\), where \(\Sigma\) is a closed \(d\)-dimensional manifold. Now there are two boundary components:
\[
    \partial M=\Sigma\times\{0\}\sqcup \Sigma\times\{1\}.
\]
In the sandwich construction \cite{Pulmann:2019vrw}, 
we put a non-topological boundary condition on \(\Sigma\times\{0\}\) and a topological boundary condition on \(\Sigma\times\{1\}\).
Now, one can then Fourier-expand the \((d+1)\)-dimensional fields satisfying the  boundary conditions to an infinite tower of \(d\)-dimensional fields. However, since the bulk is topological, all but a finite number of these fields may be integrated out, yielding a \(d\)-dimensional (generally non-topological) field theory on \(\Sigma\) with finitely many fields. Varying the topological boundary condition at \(\Sigma\times\{1\}\) leads to theories related by discrete gaugings, while varying the non-topological boundary condition at \(\Sigma\times\{0\}\) typically leads to distinct theories.

The dimensionally reduced \(d\)-dimensional action on \(\Sigma\) can be simply computed by choosing an appropriate resolution of the topological boundary condition.
The detailed method of computation can be found in \cite{Pulmann:2019vrw}, which we omit here since we do not need the details.

\section{AKSZ models as open-face sandwiches  with trivial filling}\label{sec:triv-filling}
The theories we wish to realise as sandwiches, namely the first-order formulation of Yang-Mills theory and the MacDowell-Mansouri formulation of four-dimensional general relativity, are both Manin theories. That is, non-topological perturbations of AKSZ models.
Thus, as a first step, we first realise an ordinary \(d\)-dimensional AKSZ model as a \((d+1)\)-dimensional sandwich theory.
In doing this, we will see that (1) the `filling' theory is, by itself (i.e.\ in the absence of boundaries), trivial in that the fields can be gauge-fixed to be identically zero, even in the presence of nontrivial topology; (2) only one of the two boundary conditions will be present: on the topological side, the boundary condition is effectively absent, so that we only have an `open-face sandwich' (\cref{fig:aksz_sandwich}), similar to the situation in AdS/CFT where the bulk anti-de~Sitter space only has one boundary component \cite[§6.2]{Schafer-Nameki:2023jdn}.

The fact that (2) is possible is a consequence of (1): since the `filling' is trivial, the target manifold of the bulk AKSZ model is quasi-isomorphic to a point, which has only trivial differential graded Lagrangian submanifolds.

\subsection{Weil algebra as trivial filling}
Suppose that we wish to realise the \(d\)-dimensional AKSZ model with target space given by the
differential graded symplectic manifold \((Y,d_Y,\omega_Y)\) with \(|\omega_Y|=d-1\). Let
\begin{equation}
    X = \mathrm T[1]Y
\end{equation}
be equipped with the homological vector field
\begin{equation}\label{eq:weil_differential}
    Q_X = Q_Y + v^i\frac\partial{\partial y^i},
\end{equation}
where \(y^i\) are coordinates on \(Y\) and \(v^i\) are the corresponding coordinates along the tangent directions, and \(Q_Y\) is extended to all of \(X\) via
\begin{equation}
    Q_Y\left(v^i\frac\partial{\partial y^i}f\right)+ v^i\frac\partial{\partial y^i}(Q_Yf) = 0
\end{equation}
for arbitrary \(f\in\mathcal C^\infty(X)\).
Now, on \(X\) there exists a symplectic form \(\omega_X\) of degree \(d\). Explicitly, if locally
\begin{equation}
    \omega_Y = \omega_{ij}(y)\mathrm dy^i\wedge\mathrm dy^j,
\end{equation}
then we define 
\begin{equation}
    \omega_X = \omega_{ij}(y)\mathrm dy^i\wedge\mathrm dv^j.
\end{equation}

Locally, this reduces to the Weil algebra as follows. If at \(y\in Y\) the shifted tangent space \(\mathrm T[-1]_yY\eqqcolon\mathfrak g\) is an \(L_\infty\)-algebra, then at \((y,p)\in X\) (with \(p\in\mathrm T[1]_yY\)) the shifted tangent space \(\mathrm T[-1]_{(y,p)}X\) is canonically identified with the inner automorphism \(L_\infty\)-algebra (whose associated Chevalley-Eilenberg algebra is the Weil algebra):
\begin{equation}
    \mathrm T[-1]_{(y,p)}X\cong\mathfrak{inn}(\mathfrak g).
\end{equation}
The cyclic structure of degree \(d-3\) on \(\mathfrak g\) then induces a cyclic structure of degree \(d-4\) on \(\mathfrak{inn}(\mathfrak g)\) as
\begin{equation}
    \langle y+v,y'+v'\rangle_{\mathfrak{inn}(\mathfrak g)}
    \coloneqq 
    \langle y,v'[-1]\rangle_{\mathfrak g}
    +
    \langle y',v[-1]\rangle_{\mathfrak g}
\end{equation}
for \(y,y'\in\mathfrak g\) and \(v,v'\in\mathfrak g[1]\).

Thus, on a \((d+1)\)-dimensional worldvolume \(M\), we can construct a \((d+1)\)-dimensional AKSZ model with target space \(X\), with the space of fields being
\begin{equation}
    \mathcal M\coloneqq \operatorname{Maps}(\mathrm T[1]M, X)
\end{equation}
By itself, this is a trivial \((d+1)\)-dimensional theory on the worldvolume \(M\).
To recover the \(d\)-dimensional AKSZ model with target space \(Y\),
we will need to impose appropriate boundary conditions on the boundary \(\partial M\) of \(M\), as we explain in the next section.

\subsection{Boundary conditions}
Since the target space \(X\) is (by construction) cohomologically trivial, the AKSZ theory with target space \(X\) has trivial physics without boundary conditions. Furthermore, since \(X\) is cohomologically trivial, one can (if one so chooses) \emph{not} put any boundary conditions.
Concretely, this is implemented by picking the trivial resolution
\begin{equation}\label{eq:trivial_resolution}
    l=\operatorname{id}_X\colon X\to X,
\end{equation}
which is just the identity map; this is a resolution of the (unique) trivial Lagrangian submanifold of the one-point space \(\bullet\simeq X\).

Let spacetime be \(M=\Sigma\times[0,1]\), such that the restriction of the bulk fields at one of the two boundaries is
\begin{equation}
\begin{aligned}
    \mathcal X 
    &= \operatorname{Maps}(\mathrm T[1]\Sigma,X)
    \\
    &\coloneqq\Omega(\Sigma,X[-1])
    = \mathrm T[1]\Omega(\Sigma,Y[-1]).
\end{aligned}
\end{equation}
If one puts the trivial boundary condition given by \eqref{eq:trivial_resolution} on both \(\Sigma\times\{0\}\) and \(\Sigma\times\{1\}\), then one obtains a trivial theory. Instead, let us only put the trivial boundary condition on \(\Sigma\times\{1\}\) and, where there would normally be a non-topological boundary condition, a topological (since we wish to realise a topological theory on the boundary) Dirichlet boundary condition on \(\Sigma\times\{0\}\) given by the Lagrangian submanifold \(\mathcal F\subset\mathcal X\), where
\begin{equation}
    \mathcal F = \Omega(\Sigma,Y) \subseteq \Omega(\Sigma,X).
\end{equation}
This yields a sort of `open-faced sandwich', as shown in \cref{fig:aksz_sandwich}.
Then we claim that this reduces to the ordinary AKSZ theory on \(\Sigma\).
\begin{widetext}
\begin{center}
\begin{figure}[h!]
\[
\begin{tikzpicture}[xscale=1.3, baseline=0]
    \draw[line width=5pt] (0,0) -- (4,0) node[pos=0.5, below] {\it Dirichlet boundary condition};
    \draw[line width=5pt, dashed] (0,2) -- (4,2) node[pos=.5, above] {\it trivial boundary condition};
    \draw[white, fill=gray!50, line width=0] (0,0) rectangle (4,2)
         node[pos=.5] {\color{black}\((d+1)\)-dim.\ trivial AKSZ};
\end{tikzpicture}
\overset{\text{dim.\ red.}}\implies
\begin{tikzpicture}[xscale=1.3, baseline=0]
    \draw[line width=5pt] (0,0) -- (4,0) node[pos=0.5, below] {\it nontrival \(d\)-dim.\ AKSZ};
    \draw[white, fill=white, line width=0] (0,0) rectangle (4,2);
\end{tikzpicture}
\]
\caption{
    A \(d\)-dimensional AKSZ model implemented as a \((d+1)\)-dimensional `open' sandwich theory with one trivial boundary condition (above) and one Dirichlet boundary condition (below)
}\label{fig:aksz_sandwich}
\end{figure}
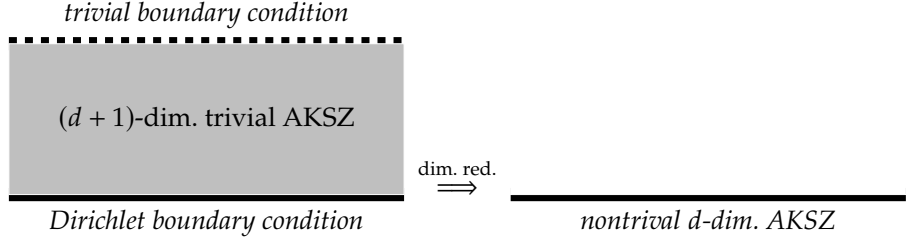
\end{center}
\end{widetext}

The absence of nontrivial topological boundary conditions at \(\Sigma\times\{1\}\) signifies the fact that this sandwich theory does \emph{not} capture any higher symmetries of the AKSZ model.

\section{Warm-up example: three-dimensional Chern-Simons theory from four dimensions}\label{sec:cs}
As a simple concrete example, let us consider the sandwich realisation of a three-dimensional Chern-Simons theory whose gauge algebra is the Lie algebra \(\mathfrak g\) with an invariant metric \(\langle~,~\rangle_{\mathfrak g}\).

\subsection{Chern-Simons theory with nontrivial bundles}
Let us review three-dimensional Chern-Simons theory as the simplest nontrivial example of an AKSZ model. If one simply takes as the target space \(\mathfrak g[1]\), then the space of fields and antifields on spacetime \(\Sigma\) is \(\Omega^\bullet(\Sigma)\otimes\mathfrak g[1]\). While this works perturbatively when \(\Sigma\) is contractible, it requires that the gauge field \(A\) be a globally defined \(\mathfrak g\)-valued one-form for instance and in particular does not allow for nontrivial \(G\)-bundles on spacetime; more importantly for our purposes, it does not remember the global topology of the Lie group \(G\) (in particular, the centre \(\operatorname Z(G)\) or the fundamental group \(\pi_1(G)\)) that gives rise to higher-form symmetry. To rectify this, we must change the target space to be a suitable \emph{stack} that remembers the global topology of \(G\).

For the purposes of Chern-Simons theory in the Batalin-Vilkovisky formalism, to a sufficiently fine open covering \(\Sigma=\bigcup_iU_i\) of spacetime, we would like to associate the following data:
\begin{subequations}\label{def:cs_data}
\begin{align}
    \mathsf A_i&\in\Omega(U_i)\otimes\mathfrak g[1],&
    g_{ij}&\in\mathcal C^\infty\mleft(U_i\cap U_j,G\mright),
\end{align}
that patch together as follows:
\begin{align}
    \mathsf A_j+\mathrm d &= g_{ij}^{-1}(\mathsf A_i+\mathrm d)g_{ij},&
    g_{ij}g_{jk}&=g_{ik},&
    g_{ii}&=1_G.
\end{align}
\end{subequations}
Thus, \(g_{ij}\) define transition maps for a principal \(G\)-bundle \(P\).
Decomposing the polyform (i.e.\ differential form of inhomogeneous degree) \(\mathsf A\) into different form degrees as \(\mathsf A=\mathsf A^{(0)}+\mathsf A^{(1)}+\mathsf A^{(2)}+\mathsf A^{(3)}\), we see that, for \(i\ne1\), \(\mathsf A^{(i)}\) transforms as differential forms valued in the associated vector bundle \(P\times_G\mathfrak g\), while \(\mathsf A^{(1)}\) transforms as a connection on \(P\). We now \emph{define} the stack \(\mathfrak g[1]/G\) to be the object that associates to an open covering the data of \eqref{def:cs_data}.

The notation \(\mathfrak g[1]/G\) is because of the following reason. In \eqref{def:cs_data}, if we had \(g_{ij}=1_G\) identically, then such objects would simply correspond to non-degree-preserving maps \(\operatorname{Maps}(\mathrm T[1]\bigcup_iU_i,\mathfrak g[1])\), so we could simply take the target space to be \(\mathfrak g[1]\). Adding the \(g_{ij}\) amounts to taking a certain quotient of this stack by an action of \(G\) \cite{Benini:2017zjv}.

Prior to quotienting, \(\mathfrak g[1]\) carries a symplectic structure of degree \(2\) given by the Killing form on \(\mathfrak g\), such that \(\Omega(U_i)\otimes\mathfrak g[1]=\operatorname{Maps}(\mathrm T[1]U_i,\mathfrak g[1])\) carries a symplectic structure of degree \(2-3=-1\). Quotienting amounts to adding in the \(g_{ij}\) and compatibility conditions, so the symplectic structure (or, rather, a Poisson structure, since the symplectic structure does not touch \(g_{ij}\)) persists on \(\operatorname{Maps}(\mathrm T[1]U_i,\mathfrak g[1]/G)\).

\subsection{Naïve sandwich construction for Chern-Simons theory}
Let us first discuss the realisation of Chern-Simons theory as a sandwich while ignoring the discrete degrees of freedom.

For Chern-Simons theory, \(Y_\mathrm{CS}=\mathfrak g[1]\) is simply (a degree-shifted) Lie algebra, where the homological vector field \(Q_X\) is given by the Chevalley-Eilenberg differential of \(\mathfrak g\) encoding the Lie bracket of \(\mathfrak g\), and the symplectic form \(\omega_{Y_\mathrm{CS}}\) of degree \(2\) on \(X\) is given by the invariant metric \(\langle~,~\rangle_{\mathfrak g}\).
Picking a basis \(t_i\) of \(\mathfrak g\), then the symplectic form and homological vector field on \(Y\) are
\begin{align}
    \omega_{Y_\mathrm{CS}} &= \omega^{ij}\mathrm dt_i \wedge\mathrm dt_j,&
    Q_{Y_\mathrm{CS}} &= \tfrac12f_i^{jk}t_jt_k\frac\partial{\partial t_i},
\end{align}
where \(\omega^{ij}\) are the structure constants for the metric on \(\mathfrak g\) and \(f^i_{jk}\) are the structure constants for the Lie bracket of \(\mathfrak g\).

Then
\begin{equation}
    X_\mathrm{CS} = \mathfrak g[2]\oplus\mathfrak g[1] = \mathfrak{inn}(\mathfrak g)[1]
\end{equation}
is the doubled symplectic differential graded manifold corresponding to the inner derivation algebra. Let us pick a basis \((\tilde t_i,t_i)\) for \(X\) such that \(\tilde t_i\) is the shifted copy of \(t_i\). Then the symplectic structure and homological vector field on \(X\) are
\begin{subequations}
\begin{align}
    \omega_X
    &= \omega^{ij} \mathrm dt_i \wedge \mathrm d\tilde t_i,
    \\
    Q_X 
    &= \tfrac12t_j t_kf_i^{jk}\frac\partial{\partial t_i}
    - t_j\tilde t_kf_i^{jk}\frac\partial{\partial\tilde t_i}
    + \tilde t_i\frac\partial{\partial t_i}.
\end{align}
\end{subequations}
The space of fields of the four-dimensional AKSZ theory on a four-manifold \(M\) with target space \(X\) is
\begin{equation}
    \operatorname{Maps}(\mathrm T[1]M,X)
    =\underbrace{\Omega(M)\otimes\mathfrak g[1]}_{\mathsf A}\times\underbrace{\Omega(M)\otimes\mathfrak g[2]}_{\mathsf B},
\end{equation}
where we have indicated the coordinates for each of the components. For instance, \(\mathsf A\) is a sum of differential forms of different degrees on \(M\); we write \(\mathsf A=\mathsf A^{(0)}+\mathsf A^{(1)}+\dotsb+\mathsf A^{(4)}\) for the decomposition into each of the form degrees.

The AKSZ action is then
\begin{equation}
    S = \int_M \mathsf B\wedge\big(\mathrm d\mathsf A+\tfrac12[\mathsf A,\mathsf A] + \mathsf B\big),
\end{equation}
where implicitly terms of form degree other than four are projected out in the integral \(\int_M\).
The non-ghostly part of this action is then
\begin{equation}
    S = \int_{M} B\wedge\big(\mathrm dA+\tfrac12[A,A] + B\big),
\end{equation}
where \(A=\mathsf A^{(1)}\) and \(B=\mathsf B^{(2)}\) are the total (form plus graded Lie algebra) degree $0$ fields in $\Omega(M)^1\otimes\mathfrak g[1]$ and $\Omega(M)^2\otimes\mathfrak g[2]$, respectively.
The equations of motion for \(A\) and \(B\) are
\begin{align}\label{eq:CSeom}
    \mathrm dB+[A,B]&=0,&
    \mathrm dA+\tfrac12[A,A]&=-\tfrac12B.
\end{align}
The BV gauge transformations are
\begin{align}
    A &\mapsto A+\mathrm dc + [A,c]-\Lambda,&
    B &\mapsto B + \mathrm d\Lambda + [A,\Lambda]
\end{align}
for the gauge parameters
\begin{align}
    c&\in\Omega^0(\Sigma\times[0,1],\mathfrak g),&
    \Lambda&\in\Omega^1(\Sigma\times[0,1],\mathfrak g)
\end{align}
with \(c=\mathsf A^{(0)}\) and \(A\in\mathsf B^{(1)}\).
If we did not have boundary conditions, then by tuning \(\Lambda\) one can gauge-fix \(A=0\), and the equations of motion then force \(B=0\), so we obtain a trivial theory as advertised.

\paragraph{Boundary conditions.}
Let \(\Sigma\) be a closed 3-manifold, and let the four-dimensional bulk spacetime be \(\Sigma\times[0,1]\).
The space of boundary fields on one of the two boundary components is
\begin{equation}
    \mathcal X = \Omega(\Sigma,X[-1])
    = \Omega(\Sigma,\mathfrak g)
    \oplus
    \Omega(\Sigma,\mathfrak g)[1].
\end{equation}
Let us label the components as \(A\in\Omega(\Sigma,\mathfrak g)\) and \(B\in \Omega(\Sigma,\mathfrak g)[1]\), with the form degrees understood to be inhomogeneous (i.e.\ \(A\) and \(B\) are polyforms).
The Dirichlet boundary condition at \(\Sigma\times\{0\}\) is
\begin{equation}
    \mathcal F = \big\{(A,B)\in\mathcal X~\big|~B=0\big\}\subset\mathcal X.
\end{equation}
That is, we are considering AKSZ fields on \(\Sigma\times[0,1]\) such that, at \(\Sigma\times\{0\}\), the \(B\) field has a Dirichlet boundary condition \(B=0\) but \(A\) is unconstrained.

With the boundary conditions imposed, we  only have gauge transformations \(\Lambda\) of the form \(\Lambda(\sigma,0)=0\). Thus, we cannot gauge away the value of \(A\) at the boundary \(\Sigma\times\{0\}\). However, we can (for example) gauge-fix
\begin{equation}
    A(\sigma,s) = A(\sigma,0)
\end{equation}
so that \(A\) is constant along \(s\). Then the equations of motion determine \(B\) in terms of \(A\), and the dynamics of \(A\) are entirely determined by its boundary value.

Since \(B\) vanishes at the boundary \(\Sigma\times\{0\}\), the equations of motion \eqref{eq:CSeom} implies $A$ is flat  along \(\Sigma\), so that we recover   three-dimensional Chern-Simons theory on \(\Sigma\), as required.

\subsection{Sandwich for centre symmetry of Chern-Simons theory}
We now modify the open sandwich to allow for discrete degrees of freedom in the four-dimensional bulk. This yields a SymTFT that captures the centre symmetry of Chern-Simons theory.

Let \(G\) be the simply connected compact Lie group with Lie algebra \(\mathfrak g\), so that its centre group \(\operatorname Z(G)=\pi_1(\tilde G)\) is the fundamental group of the centreless form \(\tilde G\coloneqq G/\operatorname Z(G)\).

In order to account for the centre symmetry, we now take the target space of the bulk four-dimensional theory to be the stack \(X_\mathrm{CS}/\tilde G\), where the action of \(\tilde G\) is now such that, if spacetime is covered by \(\Sigma=\bigcup_iU_i\) and hence \(\Sigma\times[0,1]=\bigcup_iU_i\times[0,1]\), then one has the data
\begin{align}
    (\mathsf A_i,\mathsf B_i)&\in\Omega(U_i)\otimes\mathfrak g[1]\oplus\mathfrak g[2],\\
    g_{ij}&\colon U_i\cap U_j\to G,
\end{align}
with
\begin{align}
    g_{ij}g_{jk}&\in\operatorname Z(G)g_{ik},&
    g_{ij}g_{ji}&=1_G,\\
    \mathsf A_j+\mathrm d&=g_{ij}^{-1}(\mathsf A_i+\mathrm d)g_{ij},&
    \mathsf B_j&=g_{ij}^{-1}\mathsf B_ig_{ij}.
\end{align}
Thus, \(g_{ij}\) is allowed to violate the triple overlap condition up to \(\operatorname Z(G)\) such that, effectively, the cosets \(g_{ij}\operatorname Z(G)\) define the transition maps of a principal \(\tilde G\)-bundle \(P\) over \(\Sigma\times[0,1]\). The fields \(\mathsf A\) and \(\mathsf B\) transform as differential forms valued in the associated bundle \(P\times_{\tilde G}\mathfrak g\) except for \(\mathsf A^{(1)}\), which transforms as a connection on \(P\).

Now, on a local patch, all fields can be gauge-fixed to be zero. Globally, however, the theory retains a discrete degree of freedom, namely the choice of a principal \(\tilde G\)-bundle on \(\Sigma\times[0,1]\) (or, since \(\Sigma\times[0,1]\) deformation-retracts to \(\Sigma\), a choice of a principal \(\tilde G\)-bundle on \(\Sigma\)).

The non-topological boundary condition at \(\Sigma\times\{0\}\) is as before. However, now that the \((d+1)\)-dimensional bulk theory is nontrivial, we now have more interesting choices for the topological boundary \(\Sigma\times\{1\}\). One can change the gauge group from the centreless form \(\tilde G\) to the simply connected form by imposing, at the boundary \(\Sigma\times\{1\}\),
\begin{equation}\label{eq:boundary_cond}
    g_{ij}g_{jk}=g_{ik}
\end{equation}
(at least if the large gauge transformations allow this; see note at the end of the section).
Then at the boundary \(g_{ij}\) actually define a principal \(G\)-bundle (i.e.\ the \(\tilde G\)-bundle \(P|_{\Sigma\times\{1\}}\) lifts to a \(G\)-bundle), and
the time parameter \(t\in[0,1]\) defines a homotopy of this \(G\)-bundle structure. However, since the space of topological principal \(G\)-bundles is discrete, at \(\Sigma\times\{0\}\) we must also have a lift of the \(\tilde G\)-bundle \(P|_{\Sigma\times\{1\}}\) to a \(G\)-bundle.
Thus, after dimensional reduction, the boundary condition \eqref{eq:boundary_cond} defines a \(G\)-valued Chern-Simons theory on \(\Sigma\).

More generally, given any subgroup \(\Gamma\) of \(\operatorname Z(G)\), one can impose, again assuming consistency with large gauge transformations,
\begin{equation}\label{eq:CS_gauging}
    g_{ij}g_{jk}\in\Gamma g_{ik},
\end{equation}
which yields a \(G/\Gamma\)-valued \(d\)-dimensional Chern-Simons theory on \(\Sigma\). This corresponds to gauging a \(\Gamma\) subgroup of the \(\operatorname Z(G)\)-valued electric one-form symmetry of \(G\)-valued Chern-Simons theory.

\
Note that the above discussion has assumed that in fact \eqref{eq:CS_gauging} is consistent with the large gauge transformations of the AKSZ model. This is not always the case; for instance, for \(G\) compact and for certain levels, demanding invariance under large gauge transformations implies that \eqref{eq:CS_gauging} requires \(g_{ij}\) to be trivial, as discussed in \cite{Gaiotto:2014kfa}.

\section{Symmetry topological field theory for the centre symmetry of Yang-Mills}\label{sec:ym}

\newpage

\begin{widetext}
\begin{center}
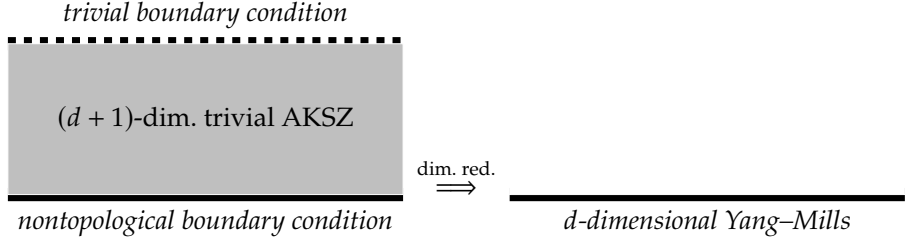
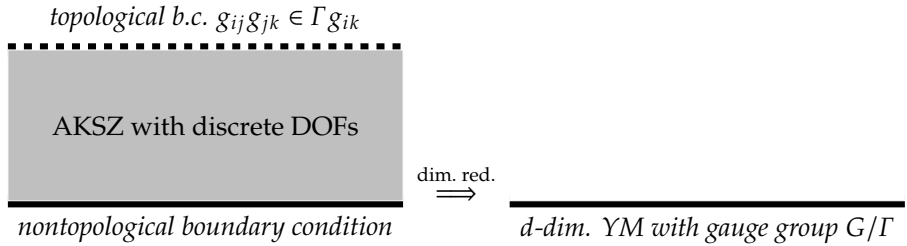
\begin{figure}[h!]
\begin{subfigure}\textwidth
\[
\begin{tikzpicture}[xscale=1.3, baseline=0]
    \draw[line width=5pt] (0,0) -- (4,0) node[pos=0.5, below] {\it nontopological boundary condition};
    \draw[line width=5pt, dashed] (0,2) -- (4,2) node[pos=.5, above] {\it trivial boundary condition};
    \draw[white, fill=gray!50, line width=0] (0,0) rectangle (4,2)
         node[pos=.5] {\color{black}\((d+1)\)-dim.\ trivial AKSZ};
\end{tikzpicture}
\overset{\text{dim.\ red.}}\implies
\begin{tikzpicture}[xscale=1.3, baseline=0]
    \draw[line width=5pt] (0,0) -- (4,0) node[pos=0.5, below] {\it \(d\)-dimensional Yang--Mills};
    \draw[white, fill=white, line width=0] (0,0) rectangle (4,2);
\end{tikzpicture}
\]
\caption{A naïve open-sandwich construction}\label{subfig:naive_ym_sandwich}
\end{subfigure}
\begin{subfigure}\textwidth
\[
\begin{tikzpicture}[xscale=1.3, baseline=0]
    \draw[line width=5pt] (0,0) -- (4,0) node[pos=0.5, below] {\it nontopological boundary condition};
    \draw[line width=5pt, dashed] (0,2) -- (4,2) node[pos=.5, above] {\it topological b.c.\ \(g_{ij}g_{jk}\in\Gamma g_{ik}\)};
    \draw[white, fill=gray!50, line width=0] (0,0) rectangle (4,2)
         node[pos=.5] {\color{black}AKSZ with discrete DOFs};
\end{tikzpicture}
\overset{\text{dim.\ red.}}\implies
\begin{tikzpicture}[xscale=1.3, baseline=0]
    \draw[line width=5pt] (0,0) -- (4,0) node[pos=0.5, below] {\it \(d\)-dim. YM with gauge group \(G/\Gamma\)};
    \draw[white, fill=white, line width=0] (0,0) rectangle (4,2);
\end{tikzpicture}
\]
\caption{Yang-Mills sandwich construction for the centre symmetry}\label{subfig:better_ym_sandwich}
\end{subfigure}
\captionsetup{subrefformat=parens}
\caption{
    A naïve open-sandwich construction \subref{subfig:naive_ym_sandwich} of Yang-Mills theory with one trivial boundary condition (above) and one Dirichlet boundary condition (below) does not capture the electric centre symmetry and does not allow for nontrivial principal bundles.
    A more refined construction \subref{subfig:better_ym_sandwich} with a nontrivial filling does capture the centre symmetry, and subgroups of the centre symmetry can be gauged by putting different topological boundary conditions (above).
}\label{fig:ym_sandwich}
\end{figure}
\end{center}
\end{widetext}

\subsection{Naïve open sandwich for Yang-Mills theory}\label{ssec:ym_naive}
An open-sandwich construction for pure Yang-Mills theory is given in \cite[§10.4]{Pulmann:2019vrw} (see \cref{subfig:naive_ym_sandwich}). This construction has a trivial filling (i.e.\ the AKSZ theory in the bulk is trivial), and hence there are no nontrivial topological boundary conditions (i.e.\ this is an open sandwich), so this construction by itself does not capture any higher symmetries of Yang-Mills theory, such as the centre symmetry. Nevertheless, it serves as a convenient base to start from, so we review this construction.

A first-order formulation of \(d\)-dimensional Yang-Mills theory with gauge group \(G\) (with Lie algebra \(\mathfrak g\) and invariant metric \(\langle-,-\rangle_{\mathfrak g}\)) is given by the action
\begin{equation}
    S = \int_\Sigma B\wedge\big(\mathrm dA +{\tfrac12}[A,A]\big) + \tfrac12 B\wedge\star B,
\end{equation}
where \(B\) is an auxiliary \(\mathfrak g^*\)-valued two-form on spacetime \(\Sigma\); upon integrating out \(B\), one recovers the familiar second-order action \(S=\int\frac12F\wedge\star F\).
This is a Manin theory \cite{Arvanitakis:2024dbu,Borsten:2024pfz,Borsten:2024alh,Arvanitakis:2025nyy}, i.e.\ an AKSZ model (in this case, a \(BF\) model) perturbed by a non-topological quadratic term; it can be realised as an open sandwich similar to the pure AKSZ model but with a modified non-topological boundary condition.

The \(d\)-dimensional \(BF\) model is given by the gauge \(L_\infty\)-algebra \(\mathfrak g\ltimes\mathfrak g^*[d-3]\), which is a graded Lie algebra (the differential vanishes) whose Lie bracket is given by the coadjoint representation of \(\mathfrak g\); it comes with the obvious canonical cyclic structure of degree \(d-3\). This corresponds to the symplectic differential graded manifold
\begin{equation}
    Y_\mathrm{YM} = \left(\mathfrak g\ltimes\mathfrak g^*[d-3]\right)[1],
\end{equation}
whose underlying graded vector space is simply \(\mathfrak g[1]\oplus\mathfrak g^*[d-2]\). Its double is then
\begin{equation}
    X_\mathrm{YM} = \mathrm T[1]Y_\mathrm{YM},
\end{equation}
whose underlying graded vector space is \(\mathfrak g[1]\oplus\mathfrak g[2]\oplus\mathfrak g^*[d-2]\oplus\mathfrak g^*[d-1]\).

We then consider the \((d+1)\)-dimensional AKSZ model with target space \(X_{\mathrm{YM}}\) and \((d+1)\)-dimensional worldvolume \(\Sigma\times[0,1]\), so that the space of fields at one of the two boundary components is
\begin{equation}
\begin{aligned}
    \mathcal X
    = \Big(
    &\overset{\mathsf A}{\Omega(\Sigma,\mathfrak g)}
    \oplus
    \overset{\tilde{\mathsf A}}{\Omega(\Sigma,\mathfrak g)[1]}
    \\
    &\oplus 
    \underset{\mathsf B}{\Omega(\Sigma,\mathfrak g^*)[d-3]}
    \oplus
    \underset{\tilde{\mathsf B}}{\Omega(\Sigma,\mathfrak g^*)[d-2]}\Big)[1],
\end{aligned}
\end{equation}
where we have indicated the coordinates for each of the components. So, for instance, \(\tilde{\mathsf A}\) is a sum of differential forms of different degrees on \(\Sigma\); we write \(\tilde{\mathsf A}=\tilde{\mathsf A}^{(0)}+\tilde{\mathsf A}^{(1)}+\dotsb+\tilde{\mathsf A}^{(d)}\) for the decomposition into each of the form degrees.

At \(\Sigma\times\{1\}\) we put no boundary conditions; at \(\Sigma\times\{0\}\), we put the boundary condition given by
\begin{equation}
	\mathcal F
	= \left\{
	\begin{pmatrix}
		\mathsf{A} \\ \tilde{\mathsf A} \\ \mathsf B \\ \tilde{\mathsf B}
	\end{pmatrix}
	\in\mathcal X
	~\middle|~
	\begin{aligned}
		&\mathsf B^{(0)}=\dotsb=\mathsf B^{(d-3)} = 0
		\\
		&\tilde{\mathsf A}^{(0)}=\tilde{\mathsf A}^{(1)}=0
		\\
		&\star\mathsf B^{(d-2)\sharp}=\tilde{\mathsf B}^{(2)}
	\end{aligned}
	\right\}
\end{equation}
where \((-)^\sharp\colon\mathfrak g^*\to\mathfrak g\) is the bijection induced by the invariant metric \(\langle-,-\rangle_{\mathfrak g}\) on \(\mathfrak g\), as shown in \cref{table:ym_boundary_cond}.
\begin{widetext}
\begin{center}
\begin{table}[h!]
\begin{center}
\begin{tabular}{lcccccccc}\toprule
&&0-form&1-form&2-form&\(\cdots\)&\((d-2)\)&\((d-1)\)&\(d\)-form\\\midrule
\(\mathfrak g[1]\)&\(\mathsf A\)&\(c\)&\(A\)&\(B^+\)&\multirow2*{\(\cdots\)}&\multicolumn3c{\multirow2*}\\
\(\mathfrak g[2]\)&\(\tilde{\mathsf A}\)&\black2&\grey{\(\star B^\sharp\)}\\
\(\mathfrak g^*[d-2]\)&\(\mathsf B\)&\black3&\blackwith{\(\cdots\)}&\grey{\(B\)}&\(A^+\)&\(c^+\)\\
\(\mathfrak g^*[d-1]\)&\(\tilde{\mathsf B}\)&\black3&\blackwith{\(\cdots\)}&\black3\\
\bottomrule
\end{tabular}
\end{center}
\caption{
    The non-topological boundary condition \(\mathcal F\) for realising Yang-Mills theory. Cells coloured black are constrained to vanish at the boundary \(\Sigma\times\{0\}\) (Dirichlet condition); cells coloured white are unconstrained; cells coloured grey have a nontrivial constraint.
}\label{table:ym_boundary_cond}
\end{table}
\end{center}
\end{widetext}

Then one can identify
\begin{subequations}
\begin{align}
    c &= \mathsf A^{(0)},
    &
    A &= \mathsf A^{(1)},
    &
    B^+ &= \mathsf A^{(2)},
    \\
    B &= \mathsf B^{(d-2)},
    &
    A^+ &= \mathsf B^{(d-1)},
    &
    c^+ &= \mathsf B^{(d)}
\end{align}
\end{subequations}
as the familiar fields of the Batalin-Vilkovisky formulation of first-order Yang-Mills theory as given in e.g.\ \cite[§5.4]{Jurco:2018sby}. The field \(\tilde{\mathsf A}^{(2)}\) is constrained by the boundary condition to be \(\star B^\sharp\) and is not independent. The fields \(\mathsf A^{(3)},\dotsc,\mathsf A^{(d)}\) form trivial pairs with \(\tilde{\mathsf A}^{(3)},\dotsc,\tilde{\mathsf A}^{(d)}\) respectively and may be trivially integrated out.\footnote{Note that the \(L_\infty\)-algebra \(\mathcal F[-1]\) does \emph{not} have a cyclic structure in the strict sense because the trivial pairs do not have corresponding antifields; after eliminating the trivial pairs one obtains the usual cyclic \(L_\infty\)-algebra encoding the Batalin-Vilkovisky action as given in \cite[§5.4]{Jurco:2018sby}. That is, one only has a `cyclic structure in cohomology'.}

\subsection{A Yang-Mills centre symmetry sandwich}
We now modify the open sandwich to allow for discrete degrees of freedom in the \((d+1)\)-dimensional bulk (\cref{subfig:better_ym_sandwich}). This yields a SymTFT that captures the centre symmetry of Yang-Mills theory.

As before, let \(G\) be the simply connected compact Lie group with Lie algebra \(\mathfrak g\), so that its centre group \(\operatorname Z(G)=\pi_1(\tilde G)\) is the fundamental group of the centreless form \(\tilde G\coloneqq G/\operatorname Z(G)\).

In order to account for the centre symmetry, we now take the target space of the bulk \((d+1)\)-dimensional theory to be the stack \(X_\mathrm{YM}/\tilde G\). The action of \(\tilde G\) is now such that, if spacetime is covered by \(\Sigma=\bigcup_iU_i\) and hence \(\Sigma\times[0,1]=\bigcup_iU_i\times[0,1]\), then one has the data
\begin{gather}
	\begin{aligned}
		(\mathsf A_i,\tilde{\mathsf A}_i,\mathsf B_i,\tilde{\mathsf B}_i)
		\in
		\Omega(U_i)
		\otimes
		\big(\mathfrak g[1]\oplus\mathfrak g[2] \oplus \mathfrak g^*[d-2]
		\\
		{}\oplus\mathfrak g^*[d-1]\big),
	\end{aligned}
    \\
    g_{ij} \colon U_i\cap U_j\to G,
\end{gather}
with
\begin{subequations}
\begin{gather}
    g_{ij}g_{jk} \in \operatorname Z(G)g_{ik},
    \\
    g_{ij}g_{ji} = 1_G,
    \\
    \mathsf A_j+\mathrm d = g_{ij}^{-1}(\mathsf A_i+\mathrm d)g_{ij},
    \\
    \tilde{\mathsf A}_j+\mathrm d = g_{ij}^{-1}(\tilde{\mathsf A_i}+\mathrm d)g_{ij},
    \\
    \mathsf B_j = g_{ij}^{-1}\mathsf B_ig_{ij},
    \\
    \tilde{\mathsf B}_j = g_{ij}^{-1}\tilde{\mathsf B_i}g_{ij}.
\end{gather}
\end{subequations}
Thus, \(g_{ij}\) is allowed to violate the triple overlap condition up to \(\operatorname Z(G)\) such that, effectively, the cosets \(g_{ij}\operatorname Z(G)\) define the transition maps of a principal \(\tilde G\)-bundle \(P\) over \(\Sigma\times[0,1]\). The fields \(\mathsf A\), \(\tilde{\mathsf A}\), \(\mathsf B\), and \(\tilde{\mathsf B}\) transform as differential forms valued in the associated bundle \(P\times_{\tilde G}\mathfrak g\) except for \(\mathsf A^{(1)}\), which transforms as a connection on \(P\).
The \((d+1)\)-dimensional AKSZ action is then
\begin{equation}
\begin{aligned}
    S_{\Sigma\times[0,1]}=\int_{\Sigma\times[0,1]}
    &\big(\mathrm d\mathsf A+\tfrac12[\mathsf A,\mathsf A]+\tilde{\mathsf A}\big)\wedge\tilde{\mathsf B}
    \\
    {}+{} &\big(\mathrm d\tilde{\mathsf A}+\tfrac12[\mathsf A,\tilde{\mathsf A}]\big)\wedge\mathsf B.
\end{aligned}
\end{equation}
On a local patch, all fields can be gauge-fixed to be zero. Globally, however, the theory retains a discrete degree of freedom, namely the choice of a principal \(\tilde G\)-bundle on \(\Sigma\times[0,1]\) (or, since \(\Sigma\times[0,1]\) deformation-retracts to \(\Sigma\), a choice of a principal \(\tilde G\)-bundle on \(\Sigma\)).

The non-topological boundary condition at \(\Sigma\times\{0\}\) is as before. However, now that the \((d+1)\)-dimensional bulk theory is nontrivial, we now have more interesting choices for the topological boundary \(\Sigma\times\{1\}\). One can change the gauge group from the centreless form \(\tilde G\) to the simply connected form by imposing, at the boundary \(\Sigma\times\{1\}\),
\begin{equation}\label{eq:boundary_cond2}
    g_{ij}g_{jk}=g_{ik}.
\end{equation}
Then at the boundary \(g_{ij}\) actually define a principal \(G\)-bundle (i.e.\ the \(\tilde G\)-bundle \(P|_{\Sigma\times\{1\}}\) lifts to a \(G\)-bundle), and
the time parameter \(t\in[0,1]\) defines a homotopy of this \(G\)-bundle structure. However, since the space of topological principal \(G\)-bundles is discrete, at \(\Sigma\times\{0\}\) we must also have a lift of the \(\tilde G\)-bundle \(P|_{\Sigma\times\{1\}}\) to a \(G\)-bundle.
Thus, after dimensional reduction, the boundary condition \eqref{eq:boundary_cond2} defines a \(G\)-valued \(d\)-dimensional Yang-Mills theory on \(\Sigma\).

More generally, given any subgroup \(\Gamma\) of \(\operatorname Z(G)\), one can impose 
\begin{equation}
    g_{ij}g_{jk}\in\Gamma g_{ik},
\end{equation}
which yields a \(G/\Gamma\)-valued \(d\)-dimensional Yang-Mills theory on \(\Sigma\). This corresponds to gauging a \(\Gamma\) subgroup of the \(\operatorname Z(G)\)-valued electric one-form symmetry of \(G\)-valued Yang-Mills theory.

\section{Symmetry topological field theory for the centre symmetry of four-dimensional gravity}\label{sec:gr}
The Pulmann-Ševera-Valach sandwich construction can easily produce theories of Manin type (i.e.\ AKSZ theory perturbed with a quadratic term). Such a formulation exists in four dimensions (but not, to the best of the authors' knowledge, in higher dimensions): the Friedel-Smolin-Starodubtsev variant \cite{Freidel:2005ak,Smolin:2003qu,Smolin:1998qp} of the MacDowell-Mansouri formulation \cite{MacDowell:1977jt,Stelle:1979va} of gravity (for reviews see \cite{Langenscheidt:2019qje,Wise:2006sm}); its properties as a Manin theory were discussed in \cite{Borsten:2024alh}.

There exists a one-form symmetry of Einstein gravity valued in the centre of the Lorentz group \(\operatorname{Spin}(1,3)\) \cite{Cheung:2024ypq}. Depending on the signature, this centre symmetry is given in \cref{table:spin_centre}.
In this section, we lift the Friedel-Smolin-Starodubtsev variant of the MacDowell-Mansouri formulation of Einstein gravity to a sandwich theory with a five-dimensional filling that provides a SymTFT that captures this \(\operatorname Z(\operatorname{Spin}(p,4-p))\) one-form centre symmetry of gravity.
\begin{widetext}
\begin{center}
\begin{table}[h!]
\begin{center}
\begin{tabular}{rccc}\toprule
spacetime signature&(0, 4)&(1, 3)&(2, 2)\\\midrule
spin group&\(\operatorname{SU}(2)\times\operatorname{SU}(2)\)&\(\operatorname{SL}(2;\mathbb C)\)&\(\operatorname{SL}(2;\mathbb R)\times\operatorname{SL}(2;\mathbb R)\)\\
centre of spin group&\(\mathbb Z_2\times\mathbb Z_2\)&\(\mathbb Z_2\)&\(\mathbb Z_2\times\mathbb Z_2\)\\
\bottomrule
\end{tabular}
\end{center}
\caption{The centre one-form symmetry group \(\operatorname Z(\operatorname{Spin}(p,4-p))\) of Einstein gravity in four dimensions with various spacetime signatures (Euclidean, Lorentzian, split). 
Here the spin group \(\operatorname{Spin}(p,4-p)\) is always taken to be connected and simply connected.}\label{table:spin_centre}
\end{table}
\end{center}
\end{widetext}

\subsection{Lightning review of the MacDowell-Mansouri formulation of gravity}
The input data of the MacDowell-Mansouri formulation is a Lie group \(G\) whose Lie algebra \(\mathfrak g\)
admits an invariant metric \(\langle-,-\rangle_{\mathfrak g}\), a closed subgroup \(H\subseteq G\) (whose Lie algebra is \(\mathfrak h\subset\mathfrak g\)), and a linear map
\begin{equation}
    \circledast\colon \mathfrak g\to\mathfrak g^*.
\end{equation}
(This map is called the `internal Hodge dual' in \cite{Wise:2006sm,Langenscheidt:2019qje}, and the notation follows that in \cite{Langenscheidt:2019qje}.)
These satisfy the following compatibility conditions:
\begin{equation}
\begin{gathered}
    \begin{aligned}
    \mathfrak g&=\mathfrak h+\ker\circledast,&
    \mathfrak h\cap\ker\circledast&=\{0\},&
    (\ker\circledast)^\flat&=\operatorname{coker}\circledast,
    \end{aligned}\\
    \circledast[h,x](y)=\circledast x([h,y])\quad\forall h\in\mathfrak h,\;x,y\in\mathfrak g,
\end{gathered}
\end{equation}
where \((-)^\flat\colon\mathfrak g\to\mathfrak g^*\) is the musical isomorphism induced by \(\langle-,-\rangle_{\mathfrak g}\). In particular, the last condition states that \(\circledast\) is \(\mathfrak h\)-equivariant.

Using this data, one can write down the action
\begin{equation}\label{eq:mm_action}
    S = \int_\Sigma B\wedge F - \tfrac12 B\wedge\circledast^\top B
\end{equation}
in terms of the fields \(A\in\Omega^1(M)\otimes\mathfrak g\) and \(B\in\Omega^2(M)\otimes\mathfrak g^*\), where \(\circledast^\top\colon\mathfrak g^*\to\mathfrak g\) is the transpose of \(\circledast\)
and where \(F=\mathrm dA+\frac12[A,A]\) is the usual field strength of \(A\).

To obtain four-dimensional Einstein gravity with a negative cosmological constant, we take \(G=\operatorname{Spin}(2,3)\) to be the anti-de~Sitter isometry group, with \(H=\operatorname{Spin}(1,3)\) the Lorentz subgroup thereof. Then we have the decomposition
\begin{equation}\label{eq:mm_decomposition}
    \mathfrak g=\mathfrak h\oplus\mathbf4,
\end{equation}
where \(\mathbf4\) is the four-vector representation of the Lorentz algebra \(\mathfrak h\). Using this, we have
\begin{equation}
    \circledast \colon t_{ij}\mapsto (3G\Lambda)\tfrac12\epsilon_{(-1)ijkl}t^{kl}
\end{equation}
for a suitable basis \(\{t_{ij}\}\) of \(\mathfrak g\) where \(i,j\in\{-1,0,1,2,3\}\) with corresponding dual basis \(\{t^{ij}\}\) of \(\mathfrak g^*\) and \(\epsilon_{ijklm}\) is the Levi-Civita symbol.
Then \(\ker\circledast = \operatorname{Span}\{t_{(-1)i}|i\in\{0,1,2,3\}\}\) and \(\operatorname{coker}\circledast = \operatorname{Span}\{t^{(-1)i}|i\in\{0,1,2,3\}\}\) are the four-vector and four-covector representations of \(\mathfrak h\). (The coefficient \(3G\Lambda\) can be matched to the cosmological constant.)

To obtain four-dimensional gravity with a positive cosmological constant, one takes \(\mathfrak g=\mathfrak o(1,4)\) to be the de~Sitter isometry algebra, with \(\mathfrak h=\mathfrak o(1,3)\) as before and \(\circledast\) defined by the decomposition \(\mathfrak g=\mathfrak h\oplus\mathbf4\).

(As an aside, let us mention some other possible choices of the data \((\mathfrak g,\mathfrak h,\circledast)\).
If \(\mathfrak h=\{0\}\) and \(\circledast=0\), then we recover the pure four-dimensional \(BF\) model \(S\propto\int_\Sigma B\wedge F\); if \(\mathfrak h=\mathfrak g\) and \(\circledast=(-)^\flat\) is the musical isomorphism, then we have \(S\propto\int_\Sigma F\wedge F\) after integrating out \(B\).)

If we had a  pure \(BF\) theory without the term \(B\wedge\circledast B\), then the gauge symmetry would be parameterised by a ghost field \(c\in\Omega^0(M)\otimes\mathfrak g\). However, the \(B\wedge\circledast B\) term breaks the gauge symmetry down to \(\ker\circledast\). That is, the complete field content of the Batalin-Vilkovisky action is (ignoring complications about nontrivial bundles)
\begin{equation}
\begin{aligned}
    c &\in \Omega^0(M)\otimes(\ker\circledast)[1],
    &
    c^+&\in\Omega^4(M)\otimes(\operatorname{coker}\circledast)[2],
    \\
    A &\in \Omega^1(M)\otimes\mathfrak g[1],
    &
    A^+ &\in \Omega^3(M)\otimes\mathfrak g^*[2],
    \\
    B &\in \Omega^2(M)\otimes\mathfrak g^*[2],
    &
    B^+ &\in \Omega^2(M)\otimes\mathfrak g[1],
\end{aligned}
\end{equation}
where \(c\) and \(c^+\) are the ghost and the ghost antifield respectively and \((-)^+\) denotes the corresponding antifield of a field.
This field content is summarised in \cref{table:mm_field_content}.
\begin{widetext}
\begin{center}
\begin{table}[h!]
\begin{center}
\begin{tabular}{lccccc}\toprule
&0-form&1-form&2-form&3-form&4-form\\\midrule
\(\mathfrak g[1]\)&\grey{\(c\)}&\(A\)&\(B^+\)&\black2\\
\(\mathfrak g^*[2]\)&\black2&\(B\)&\(A^+\)&\grey{\(c^+\)}\\
\bottomrule
\end{tabular}
\end{center}
\caption{Field content of the MacDowell-Mansouri formulation of Einstein gravity.
The blacked out cells correspond to those fields of the Batalin-Vilkovisky formulation of the \(BF\) model that are not present in the MacDowell-Mansouri formulation due to the reduced gauge symmetry.
The grey cells correspond to those fields of Batalin-Vilkovisky formulation of the \(BF\) model
that are constrained (but nonzero)  the MacDowell-Mansouri formulation due to the reduced gauge symmetry.
}\label{table:mm_field_content}
\end{table}
\end{center}
\end{widetext}

\paragraph{Reduction to gravity.}
For completeness, let us briefly indicate how the action \eqref{eq:mm_action} reduces to Einstein gravity, as already explained in \cite{Freidel:2005ak,Smolin:2003qu,Smolin:1998qp}.
Under the branching \eqref{eq:mm_decomposition}, the field \(A\) decomposes into
\[
    A=(\omega,e),
\]
where \(\omega\in\Omega^1(M)\otimes\mathfrak g\) is the spin connection and \(e\in\Omega^1(M)\otimes\mathbf4\) is the vierbein; similarly, the field strength \(F=\mathrm dA+\frac12[A,A]\) decomposes into
\[
    F
    =
    \Big(\overbrace{\mathrm d\omega+\tfrac12\omega\wedge\omega}^{\smash{\text{Riemann curvature}}}+\tfrac12 e\wedge e
    ~,~
    \overbrace{\mathrm de+\omega\wedge e}^{\text{torsion}}\Big)
\]
where the second component \(\mathrm d\omega+\frac12\omega\wedge\omega\) is the torsion of the spin connection (regarded as a vector-valued two-form), while the first term contains the Riemann curvature (regarded as an \(\mathfrak g\)-valued two-form). The structure of the \(B\wedge\circledast B\) term means that \(B\) acts as a Lagrange multiplier for the component of \(F\) in \(\ker\circledast\), namely the torsion, so that we have a torsion-free connection; assuming the invertibility of \(e\), then \(\omega\) is fixed in terms of \(e\). Integrating out \(B\) then produces the Palatini action for gravity,
\[
    S = \int\epsilon_{abcd}\Big[\big(\mathrm d\omega+\tfrac12[\omega,\omega]\big)^{ab}
    -\tfrac16\Lambda Ge^a\wedge e^b\Big]\wedge e^c\wedge e^d.
\]

In comparison to gravity, we note two possible nonperturbative issues. First, this construction does not require the vierbein \(e\) to be invertible. In the regime where \(e\) is not invertible, the theory behaves more like Yang-Mills theory \cite{Borsten:2024pfz,Borsten:2024alh}. Similarly, by default this construction allows arbitrary \(\operatorname{Spin}(1,3)\)-bundles on spacetime \(\Sigma\), whereas in gravity one often restricts this bundle to the frame bundle.
This does not affect the perturbative theory, but nonperturbatively one may wish to include nontrivial bundles \cite{Borsten:2025phf}.

\subsection{Naïve open sandwich for gravity}

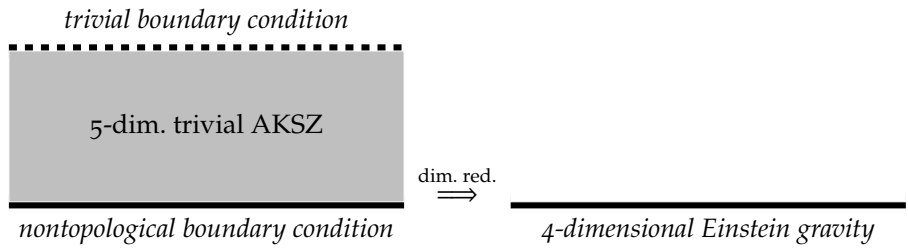
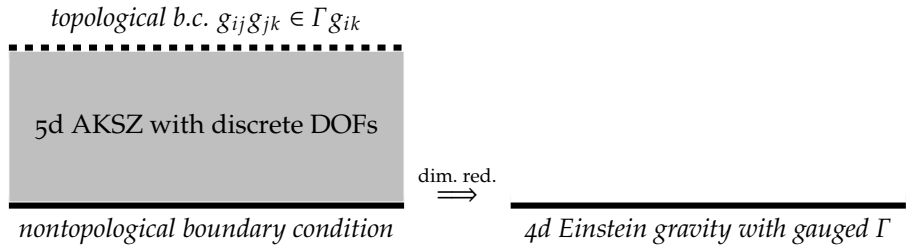
\begin{figure*}[!t]
\begin{subfigure}\textwidth
\[
\begin{tikzpicture}[xscale=1.3, baseline=0]
    \draw[line width=5pt] (0,0) -- (4,0) node[pos=0.5, below] {\it nontopological boundary condition};
    \draw[line width=5pt, dashed] (0,2) -- (4,2) node[pos=.5, above] {\it trivial boundary condition};
    \draw[white, fill=gray!50, line width=0] (0,0) rectangle (4,2)
         node[pos=.5] {\color{black}5-dim.\ trivial AKSZ};
\end{tikzpicture}
\overset{\text{dim.\ red.}}\implies
\begin{tikzpicture}[xscale=1.3, baseline=0]
    \draw[line width=5pt] (0,0) -- (4,0) node[pos=0.5, below] {\it 4-dimensional Einstein gravity};
    \draw[white, fill=white, line width=0] (0,0) rectangle (4,2);
\end{tikzpicture}
\]
\caption{A naïve open-sandwich construction for gravity}\label{subfig:naive_gr_sandwich}
\end{subfigure}
\begin{subfigure}\textwidth
\[
\begin{tikzpicture}[xscale=1.3, baseline=0]
    \draw[line width=5pt] (0,0) -- (4,0) node[pos=0.5, below] {\it nontopological boundary condition};
    \draw[line width=5pt, dashed] (0,2) -- (4,2) node[pos=.5, above] {\it topological b.c.\ \(g_{ij}g_{jk}\in\Gamma g_{ik}\)};
    \draw[white, fill=gray!50, line width=0] (0,0) rectangle (4,2)
         node[pos=.5] {\color{black}5d AKSZ with discrete DOFs};
\end{tikzpicture}
\overset{\text{dim.\ red.}}\implies
\begin{tikzpicture}[xscale=1.3, baseline=0]
    \draw[line width=5pt] (0,0) -- (4,0) node[pos=0.5, below] {\it 4d Einstein gravity with gauged \(\Gamma\)};
    \draw[white, fill=white, line width=0] (0,0) rectangle (4,2);
\end{tikzpicture}
\]
\caption{Sandwich construction for gravity capturing the centre symmetry}\label{subfig:better_gr_sandwich}
\end{subfigure}
\captionsetup{subrefformat=parens}
\caption{
    A naïve open-sandwich construction \subref{subfig:naive_gr_sandwich} of the MacDowell-Mansouri formulation of Einstein gravity with one trivial boundary condition (above) and one Dirichlet boundary condition (below) does not capture the electric centre symmetry and does not allow for nontrivial principal bundles.
    A more refined construction \subref{subfig:better_gr_sandwich} with a nontrivial filling does capture the centre symmetry of gravity, which can be gauged by putting different topological boundary conditions (above).
}\label{fig:gr_sandwich}
\end{figure*}

First, we present the open-sandwich construction (\cref{subfig:naive_gr_sandwich}) for the MacDowell-Mansouri formulation of gravity
ignoring the nontrivial \(H\)-bundle structure as in \cref{ssec:ym_naive}. This produces an open sandwich with a five-dimensional filling that is trivial and does not capture any symmetries.

The target space of the four-dimensional \(BF\) model without the \(B\wedge\circledast B\) term is the symplectic differential graded manifold
\begin{equation}
    Y_\mathrm{GR} = (\mathfrak g\ltimes\mathfrak g^*[1])[1]
\end{equation}
with the homological vector field \(Q\) defined by the graded Lie algebra structure on \(\mathfrak g\ltimes\mathfrak g^*[1]\) (a semidirect sum of the Lie algebra \(\mathfrak g\) and the Abelian graded Lie algebra \(\mathfrak g^*[1]\) using the coadjoint action of the former on the latter) and a symplectic structure of intrinsic degree \(3\) (given by the identification \((\mathfrak g\ltimes\mathfrak g^*[1])[1]\cong\mathrm T^*[3](\mathfrak g[1])\) as a cotangent bundle). The five-dimensional filling theory is the AKSZ model with target space given by the double
\begin{equation}
    X = \mathrm T[1]Y
    \cong \mathfrak g[1]\times \mathfrak g[2]\times\mathfrak g^*[2]\times\mathfrak g^*[3].
\end{equation}
This has trivial cohomology, so the five-dimensional filling theory is trivial if no boundary conditions are imposed.

The space of boundary fields at \(\Sigma\times\{0\}\subset\Sigma\times[0,1]\) is then
\begin{equation}
\begin{aligned}
	\mathcal X 
	={} &\overset{\mathsf{A}}{\Omega(\Sigma,\mathfrak{g})[1]}
	\oplus \overset{\mathsf{\tilde{\mathsf{A}}}}{\Omega(\Sigma,\mathfrak g)[2]}
	\\
	&\oplus \underset{\mathsf{B}}{\Omega(\Sigma,\mathfrak{g}^\ast)[2]}
	\oplus \underset{\mathsf{\tilde B}}{\Omega(\Sigma,\mathfrak{g}^\ast)[3]}
\end{aligned}
\end{equation}
where we have indicated polyform coordinates \(\mathsf A,\tilde{\mathsf A},\mathsf B,\tilde{\mathsf B}\) for each component.
On the topological end \(\Sigma\times\{1\}\), we impose no boundary conditions (i.e.\ we have an open sandwich).
On the non-topological end \(\Sigma\times\{0\}\), we impose the boundary condition given by the Lagrangian submanifold
\begin{equation}
	\mathcal{F}
	=
	\left\{
	\begin{pmatrix}
		\mathsf A \\ \tilde{\mathsf A} \\ \mathsf B \\ \tilde{\mathsf B}
	\end{pmatrix}
	\in\mathcal X
	~\middle|~
	\begin{aligned}
		&\tilde{\mathsf B} = 0
		\\
		&\tilde{\mathsf A}^{(0)}= \tilde{\mathsf A}^{(1)} = 0
		\\
		&\mathsf B^{(0)} = \mathsf B^{(1)} = 0
    		\\
    		&\circledast\tilde{\mathsf A}^{(2)}=\mathsf B^{(2)}
    		\\
    		&\mathsf A^{(0)}\in\Omega^0(\Sigma)\otimes(\ker\circledast)
    		\\
    		&\mathsf B^{(4)}\in\Omega^0(\Sigma)\otimes(\operatorname{coker}\circledast)
	\end{aligned}
	\right\}
\end{equation}
as shown in \eqref{fig:gr_boundary_cond}.
\begin{table}[h!]
\begin{center}
\begin{tabular}{lcccccc}\toprule
&&0-form&1-form&2-form&3-form&4-form\\\midrule
\(\mathfrak g[1]\)&\(\mathsf A\)&\grey{\(c\)}&\(A\)&\(B^+\)&\multicolumn2c{\multirow2*{(trivial pairs)}}\\
\(\mathfrak g[2]\)&\(\tilde{\mathsf A}\)&\black2&\grey{\(\circledast B\)}\\
\(\mathfrak g^*[2]\)&\(\mathsf B\)&\black2&\(B\)&\(A^+\)&\grey{\(c^+\)}\\
\(\mathfrak g^*[3]\)&\(\tilde{\mathsf B}\)&\black5\\
\bottomrule
\end{tabular}
\end{center}
\caption{Non-topological boundary condition for the sandwich realisation of the MacDowell-Mansouri of Einstein gravity.
Cells coloured in black are constrained to vanish at the boundary (Dirichlet condition);
cells coloured in grey are constrained in more complicated ways.
After dimensional reduction, the cells labelled `(trivial pairs)' correspond to trivial pairs that can be eliminated.}\label{fig:gr_boundary_cond}
\end{table}

After dimensional reduction, then, it is straightforward to see that one obtains the action \eqref{eq:mm_action} together with ghosts.

\subsection{Sandwich for centre symmetry of gravity}
We now modify the open sandwich to allow for discrete degrees of freedom in the five-dimensional bulk (\cref{subfig:better_gr_sandwich}). This yields a SymTFT that captures the centre symmetry of gravity.

Let \(\tilde H\) be the centreless form \(\operatorname{PSO}(p,4-p)\) of the simply connected Lorentz group \(H=\operatorname{Spin}(p,4-p)\).

In order to account for the centre symmetry, we now take the target space of the bulk \((d+1)\)-dimensional theory to be the stack \(X_\mathrm{GR}/\tilde H\). The action of \(\tilde H\) is such that, if spacetime is covered by \(\Sigma=\bigcup_iU_i\) and hence \(\Sigma\times[0,1]=\bigcup_iU_i\times[0,1]\), then one has the data
\begin{gather}
	\begin{aligned}
    		(\mathsf A_i,\tilde{\mathsf A}_i,\mathsf B_i,\tilde{\mathsf B}_i)
    		\in
    		\Omega(U_i)\otimes
    		\big(&\mathfrak g[1]\oplus\mathfrak g[2]
    		\\
    		&{}\oplus\mathfrak g^*[2]\oplus\mathfrak g^*[3]\big),
    \end{aligned}
    \\
    g_{ij} \colon U_i\cap U_j\to H,
\end{gather}
with
\begin{subequations}
\begin{gather}
    g_{ij}g_{jk} \in \operatorname Z(H)g_{ik},
    \\
    g_{ij}g_{ji} = 1_H,
    \\
    \mathsf A_j+\mathrm d = g_{ij}^{-1}(\mathsf A_i+\mathrm d)g_{ij},
    \\
    \tilde{\mathsf A}_j+\mathrm d = g_{ij}^{-1}(\tilde{\mathsf A_i}+\mathrm d)g_{ij},
    \\
    \mathsf B_j = g_{ij}^{-1}\mathsf B_ig_{ij},
    \\
    \tilde{\mathsf B}_j = g_{ij}^{-1}\tilde{\mathsf B_i}g_{ij}.
\end{gather}
\end{subequations}
Thus, \(g_{ij}\) is constrained to lie in \(H\) (not \(G\)) and allowed to violate the triple overlap condition up to \(\operatorname Z(H)\) such that, effectively, the cosets \(g_{ij}\operatorname Z(H)\) define the transition maps of a principal \(\tilde H\)-bundle \(P_{\tilde H}\) over \(\Sigma\times[0,1]\), which trivially extends to a principal \(\tilde G\)-bundle \(P_{\tilde G}\) over \(\Sigma\times[0,1]\).
The fields \(\mathsf A\), \(\tilde{\mathsf A}\), \(\mathsf B\), and \(\tilde{\mathsf B}\) transform as differential forms valued in the associated bundle \(P_{\tilde G}\times_{\tilde G}\mathfrak g\) except for \(\mathsf A^{(1)}\), which transforms as a connection on \(P_{\tilde G}\).
The \((d+1)\)-dimensional AKSZ action is then
\begin{equation}
\begin{aligned}
    S_{\Sigma\times[0,1]} = \int_{\Sigma\times[0,1]}
    &\big(\mathrm d\mathsf A + \tfrac12 [\mathsf A,\mathsf A]+\tilde{\mathsf A}\big)\wedge\tilde{\mathsf B}
    \\
    {}+{} & \big(\mathrm d\tilde{\mathsf A}+\frac12[\mathsf A,\tilde{\mathsf A}]\big)\wedge\mathsf B.
\end{aligned}
\end{equation}
On a local patch, all fields can be gauge-fixed to be zero. Globally, however, the theory retains a discrete degree of freedom, namely the choice of a principal \(\tilde H\)-bundle on \(\Sigma\times[0,1]\) (or, since \(\Sigma\times[0,1]\) deformation-retracts to \(\Sigma\), a choice of a principal \(\tilde H\)-bundle on \(\Sigma\)).

The non-topological boundary condition at \(\Sigma\times\{0\}\) is as before. However, now that the \((d+1)\)-dimensional bulk theory is nontrivial, we now have more interesting choices for the topological boundary \(\Sigma\times\{1\}\). One can change the gauge group from the centreless form \(\tilde G\) to the simply connected form by imposing, at the boundary \(\Sigma\times\{1\}\),
\begin{equation}\label{eq:boundary_cond_gr}
    g_{ij}g_{jk}=g_{ik}.
\end{equation}
Then at the boundary \(g_{ij}\) actually define a principal \(H\)-bundle (i.e.\ the \(\tilde G\)-bundle \(P|_{\Sigma\times\{1\}}\) lifts to a \(H\)-bundle), and
the time parameter \(t\in[0,1]\) defines a homotopy of this \(G\)-bundle structure. However, since the space of topological principal \(G\)-bundles is discrete, at \(\Sigma\times\{0\}\) we must also have a lift of the \(\tilde H\)-bundle \(P|_{\Sigma\times\{1\}}\) to a \(H\)-bundle (that is, a spin structure).
Thus, after dimensional reduction, the boundary condition \eqref{eq:boundary_cond_gr} defines ordinary Einstein gravity on \(\Sigma\) with a spin structure.

More generally, given any subgroup \(\Gamma\) of \(\operatorname Z(H)\), one can impose 
\begin{equation}
    g_{ij}g_{jk}\in\Gamma g_{ik}.
\end{equation}
This corresponds to gauging a \(\Gamma\) subgroup of the \(\operatorname Z(H)\)-valued electric one-form symmetry of Einstein gravity theory.

\section*{Acknowledgements}
L.B. and D.K. thank Sakura Schäfer-Nameki\textsuperscript{\orcidlink{0000-0003-0138-0407}} for helpful conversations that initiated this project. L.B. thanks Mathematical Institute, University of Oxford for an invitation to give a seminar and their hospitality, which proved instrumental for the writing of this paper.

\section*{Note added in proof}
While this paper was in its final stages of editing, the paper \cite{Apruzzi:2025hvs} appeared, which  addresses the same topic of SymTFTs for gravity (with  a complementary  framework and focus).

\bibliography{biblio}

\end{document}